\documentclass[dvips]{acta}
\usepackage{supertabular,lscape,epsfig}
\usepackage{amssymb}
\usepackage{amsmath}
\DeclareSymbolFont{ppa}{OT1}{ppl}{m}{it}
\DeclareMathSymbol{\vv}{\mathalpha}{ppa}{'166}

\newfont{\hb}{rphvb at 10pt}
\newfont{\hbo}{rphvbo at 10pt}
\newfont{\bitt}{rptmbi at 12pt}
\newfont{\bits}{rptmbi at 11pt}

\SetPages{239}{266}

\SetVol{63}{2013}

\begin{document}

\newcommand{\TabApp}[2]{\begin{center}\parbox[t]{#1}{\centerline{
  {\bf Appendix}}
  \vskip2mm
  \centerline{\small {\spaceskip 2pt plus 1pt minus 1pt T a b l e}
  \refstepcounter{table}\thetable}
  \vskip2mm
  \centerline{\footnotesize #2}}
  \vskip3mm
\end{center}}

\newcommand{\TabCapp}[2]{\begin{center}\parbox[t]{#1}{\centerline{
  \small {\spaceskip 2pt plus 1pt minus 1pt T a b l e}
  \refstepcounter{table}\thetable}
  \vskip2mm
  \centerline{\footnotesize #2}}
  \vskip3mm
\end{center}}

\newcommand{\TTabCap}[3]{\begin{center}\parbox[t]{#1}{\centerline{
  \small {\spaceskip 2pt plus 1pt minus 1pt T a b l e}
  \refstepcounter{table}\thetable}
  \vskip2mm
  \centerline{\footnotesize #2}
  \centerline{\footnotesize #3}}
  \vskip1mm
\end{center}}

\newcommand{\MakeTableApp}[4]{\begin{table}[p]\TabApp{#2}{#3}
  \begin{center} \TableFont \begin{tabular}{#1} #4 
  \end{tabular}\end{center}\end{table}}

\newcommand{\MakeTableSepp}[4]{\begin{table}[p]\TabCapp{#2}{#3}
  \begin{center} \TableFont \begin{tabular}{#1} #4 
  \end{tabular}\end{center}\end{table}}

\newcommand{\MakeTableee}[4]{\begin{table}[htb]\TabCapp{#2}{#3}
  \begin{center} \TableFont \begin{tabular}{#1} #4
  \end{tabular}\end{center}\end{table}}

\newcommand{\MakeTablee}[5]{\begin{table}[htb]\TTabCap{#2}{#3}{#4}
  \begin{center} \TableFont \begin{tabular}{#1} #5 
  \end{tabular}\end{center}\end{table}}

\newfont{\bb}{ptmbi8t at 12pt}
\newfont{\bbb}{cmbxti10}
\newfont{\bbbb}{cmbxti10 at 9pt}
\newcommand{\uprule}{\rule{0pt}{2.5ex}}
\newcommand{\douprule}{\rule[-2ex]{0pt}{4.5ex}}
\newcommand{\dorule}{\rule[-2ex]{0pt}{2ex}}
\def\thefootnote{\fnsymbol{footnote}}
\begin{Titlepage}
\Title{Evolution of Cool Close Binaries -- Rapid Mass Transfer\\ 
and Near Contact Binaries}
\vspace*{7pt}
\Author{K.~~ S~t~ê~p~i~e~ñ~~ and~~ M.~~ K~i~r~a~g~a}{
Warsaw University Observatory, Al.~Ujazdowskie~4, 00-478~Warsaw, Poland\\
email: (kst,kiraga)@astrouw.edu.pl}
\vspace*{7pt}
\Received{May 21, 2013}
\end{Titlepage}

\vspace*{5pt}
\Abstract{We test the evolutionary model of cool close
binaries developed by one of us (KS) on the observed properties of near
contact binaries (NCBs). These are binaries with one component filling the
inner critical Roche lobe and the other almost filling it. Those with a
more massive component filling the Roche lobe are SD1 binaries whereas in
SD2 binaries the Roche lobe filling component is less massive. Our
evolutionary model assumes that, following the Roche lobe overflow by the
more massive component (donor), mass transfer occurs until mass ratio
reversal. A binary in an initial phase of mass transfer, before mass
equalization, is identified with SD1 binary. We show that the transferred
mass forms an equatorial bulge around the less massive component
(accretor). Its presence slows down the mass transfer rate to the value
determined by the thermal time scale of the accretor, once the bulge sticks
out above the Roche lobe. It means, that in a binary with a (typical) mass
ratio of 0.5 the SD1 phase lasts at least 10 times longer than resulting
from the standard evolutionary computations neglecting this effect. This is
why we observe so many SD1 binaries. Our explanation is in contradiction to
predictions identifying the SD1 phase with a broken contact phase of the
Thermal Relaxation Oscillations model. The continued mass transfer, past
mass equalization, results in mass ratio reversed. SD2 binaries are
identified with this phase. Our model predicts that the time scales of SD1
and SD2 phases are comparable to one another.

Analysis of the observations of 22 SD1 binaries, 27 SD2 binaries and 110
contact binaries (CBs) shows that relative number of both types of NCBs
favors similar time scales of both phases of mass transfer. Total masses,
orbital angular momenta and orbital periods of SD1 and SD2 binaries are
indistinguishable from each other whereas they differ substantially from
the corresponding parameters of CBs. We conclude that the results of the
analysis fully support the model presented in this paper.}
{Stars: activity -- binaries: close -- Stars:
evolution -- Stars: late-type -- Stars: rotation}

\Section{Introduction}
Near contact binaries (NCBs) are a class of cool close binaries showing
eclipses, with one component filling the inner critical equipotential
surface (Roche lobe) and the other almost filling it. The name was
suggested by Shaw (1990) who dis-\break
\newpage\noindent
tinguished two types of NCBs: those with the more massive component filling
the Roche lobe were named V1010 Oph type and binaries with the less massive
component filling the Roche lobe were named FO Vir type. As Shaw states,
the massive components look in both types as normal, or ``a bit evolved''
Main Sequence (MS) stars, whereas low mass components are oversized for
their masses: $\approx1.2$ times in V1010~Oph type and 2--3 times in
FO~Vir type binaries. NCBs of V1010~Oph type show shortening of the orbital
period and O'Connell effect in their light curves, interpreted as resulting
from a hot spot situated on the trailing side of the less massive
component. NCBs of FO~Vir type never show O'Connell effect and period
lengthening prevails in them. Yakut and Eggleton (2005) suggested the name
SD1 and SD2 for NCBs of V1010 Oph and FO~Vir type, respectively. We will
use their designations in the following.

The ranges of component masses and orbital periods of NCBs overlap with the
ranges of cool contact binaries (CB), which implies a relationship between
these two kinds of binaries. The shapes of light curve differ, however,
substantially: while equal, or almost equal depth minima are observed for
CBs, the minima of NCBs are distinctly different. SD1 binaries have
sometimes been identified with the broken contact phase of the Thermal
Relaxation Oscillation (TRO) model of CBs described in detail below. The
presence of such a phase is an important constituent of this
model. Because, however, SD2 binaries do not fit to the TRO model, several
authors suggested that they may represent another route to CB formation,
resulting from mass transfer between the components with mass ratio
reversed (as in case of Algols). The evolutionary model of CBs suggested by
one of us (KS) assumes that all cool CBs are formed this way and their
components are in thermal equilibrium with no need for TROs. The main
purpose of the present investigation is to show that the observed
properties of NCBs cannot be reconciled with the TRO model but they fit
well into our model. Before we place NCBs into the evolutionary sequence
leading to CB formation, we describe first the historical development of
our understanding of origin, evolution and observational properties of cool
CBs.

Eclipsing binaries of W~UMa-type were discovered more than a century ago
(M{\"u}ller and Kempf 1903). The analysis of the light and velocity curves
carried out then showed that the binaries were very unusual, with the
components in contact and the uniform surface brightness. According to the
present definition, a cool CB, or W~UMa-type binary consists of two lower
MS stars surrounded by a common envelope lying between the inner and outer
Lagrangian zero velocity equipotential surfaces (called also the Roche
lobes) and possessing almost identical mean surface brightnesses (Mochnacki
1981). The more massive, primary component is a MS star, lying often close
to Zero Age MS (ZAMS), and the secondary component is oversized, compared
to its expected ZAMS size. The accepted upper limit is 1~d for the orbital
period and 2.5--3~\MS\ for the total mass of a cool CB although some
authors prefer somewhat higher values. The observed lower limits for the
orbital period and total mass are about 0.2~d and 1.1~\MS, respectively.

Based on early observations, several papers explaining the observed
properties of CBs were published in 50-ties and 60-ties of the past century
but no satisfactory model was obtained. Since then, an enormous amount of
new data have been collected on these stars, yet their structure and
evolutionary status still remain obscured.

Following the original suggestion by Jeans (1928) it was believed for a
long time that CBs are formed by fission of the protostellar core. An
apparently satisfactory agreement of the observed basic parameters of W~UMa
type stars with predictions resulting from the fission mechanism was
obtained by Roxburgh (1966). This strengthened the conviction that CBs are
born as contact systems with a uniform chemical composition. Yet, a
fundamental difficulty in explaining their structure remained. As Kuiper
(1941) demonstrated, the ratio of the thermal equilibrium radii for two
ZAMS stars is approximately directly proportional to the mass ratio whereas
the geometry of the inner critical Roche lobes requires that it scales as,
approximately, a square root of mass ratio. These two conditions can be
fulfilled only when both components have identical masses. Two different
zero age stars in thermal equilibrium do not fit into their Roche
lobes. This is still true for the (coeval) binary components burning
hydrogen in their cores as long as their masses are close to their initial
values. The fact that we do observe contact binaries with different
component masses, is known as the Kuiper paradox. Let us note in this
place, however, that the Kuiper paradox does not apply to evolved binaries
past mass exchange with mass ratio reversal (\eg Algols) because each
component obeys a different mass-radius relation, depending on its
evolutionary status. We will discuss this problem later.

A very convincing attempt to solve the Kuiper paradox was undertaken by
Lucy (1968). To explain the existence of two unevolved stars in contact, he
noted that the mass-radius relation for stars with proton-proton (pp)
nuclear cycle has a different slope, compared to the same relation for
stars with CNO cycle. As a result, the Kuiper paradox can be solved by
assuming that the pp cycle dominates in one component and the CNO cycle
dominates in the other one, and that convection zones of both stars share
the same adiabatic constant. This situation requires a very efficient
energy transport from primary to secondary component through the common
part of the convective envelopes. However, detailed calculations showed
that realistic models can only be produced within a narrow range of
component masses, contrary to what was observed. A model giving more
flexibility in selection of the component masses was suggested several
years later also by Lucy (1976). He still came out from the two fundamental
assumptions: first, that CBs are formed at ZAMS and second, that the
specific entropy is identical in the convection zones of both
components. To explain the Kuiper paradox for a broad range of component
masses he abandoned the assumption of thermal equilibrium for each
component separately. A similar model was concurrently developed by
Flannery (1976). The model is known as TRO model. Both components are
supposed to oscillate about the equilibrium state, with the whole binary
remaining in the global thermal equilibrium. The energy is transported from
primary to secondary {\it via} a turbulent convection which results in equal
entropies of both convective envelopes, hence equal surface
brightnesses. Lucy did not consider details of the energy transfer; he
simply assumed that on reaching contact between the components, entropies
of both convection zones equalize instantaneously. Later, TRO model was
also applied to initially detached binaries in which a primary reaches its
Roche lobe after some time spent on MS (\eg Webbink 1976, 1977ab, Sarna
and Fedorova 1989, Yakut and Eggleton 2005). Detailed modeling of such
binaries shows that, following the Roche lobe overflow (RLOF), the primary
transfers about 0.1~\MS\ to the secondary which expands and fills its Roche
lobe, forming a CB. Then, the TRO paradigm applies with a direction of
secular mass transfer reversed, \ie from the secondary to primary, until
both components merge.

TRO model explains two basic observational facts about W~UMa-type stars:
the geometry of the binary where the primary component is an ordinary MS
star and the secondary is also a MS star but swollen to the size of its
Roche lobe by energy transfer, and equal apparent effective temperatures of
both components resulting in a characteristic light curve with two equal
minima. It also gives an additional observational prediction, that a binary
oscillates between two states: contact -- when both stars fill their
respective Roche lobes and mass flows from the secondary to the primary,
and semidetached -- when the primary still fills its Roche lobe but the
secondary detaches from its lobe and mass flows from the primary to the
secondary (Lucy 1976). Time scales of both states should be comparable to
one another. Additional effects, like stellar evolution and/or angular
momentum loss (AML) can influence the ratio of both time scales reducing
the duration of the semidetached state (Robertson and Eggleton 1977,
Rahunen 1983, Yakut and Eggleton 2005, Li, Han and Zhang 2005). In the most
extreme case a binary can remain in contact all the time if AML rate is
high enough but then its lifetime as a CB must be as short as
$\approx10^8$~y as noted by Webbink (2003). Such binaries would be rare
in space, contrary to observations.

As more and more theoretical and observational data on CBs are accumulated,
the TRO model encounters increasing difficulties. In particular, its both
basic assumptions, \ie zero age of CBs and identical specific entropy in
the convection zones of both components, are questionable and its basic
prediction of the broken contact phase seems to be at odds with
observations.

Fission of a contracting protostellar core has been abandoned many years
ago and it is no longer considered a feasible mechanism for binary
formation. Numerical simulations of the bar-like instability developing in
a rapidly rotating liquid body showed that it never results in a fission if
the liquid is compressible (which a star is). Instead, a spiral arm
structure develops resulting in a disk, or ring containing the excess of
angular momentum (AM) and a stable core (Bonnell 2001). At present, the
early fragmentation of a protostellar cloud resulting in two protostellar
cores orbiting each other is considered to be a dominating mechanism for
binary formation (Machida \etal 2008 and references herein). In effect, the
orbit size of a freshly formed binary must be large enough to accommodate
both pre-MS components. In particular, the low period limit for the
isolated progenitors of W UMa-type binaries with total masses between
1.5~\MS\ and 3~\MS\ is about 1.5--2.0~d. Recently, a mechanism for
tightening of the originally wide, eccentric orbit by an interaction with a
properly placed, distant third companion has been suggested. The
interaction enforces the, so called, Kozai cycles which, together with the
tidal friction (KCTF), make the period of the inner binary
shorten. Detailed calculations showed that the inner period can be
shortened down to a value of about 2--3~d only (Eggleton and
Kiseleva-Eggleton 2006, Fabrycky and Tremaine 2007). The orbit is
circularized at this value and the mechanism does not work any more. So,
excluding very exceptional cases, we do not expect the KCTF mechanism to
produce binaries with periods shorter than this limit. The typical time
scale of the period shortening of a binary with the initial period of
10--20~d is of the order of $10^6{-}10^7$~y. We should then observe an
excess of young binaries with periods of 2--3~d compared to the
``canonical'' Duquenoy and Mayor (1991) distribution. This is indeed
observed among binaries in Hyades (Griffin 1985, Stêpieñ 1995). Tokovinin
\etal (2006) demonstrated, on the other hand, that almost all field
binaries with periods shorter than~3 d are in fact triple systems, compared
to only about one third of long period binaries possessing a tertiary
companion. This suggests that the KCTF mechanism works indeed efficiently,
producing large numbers of short period binaries compared to those formed
as isolated systems.

The observations of pre-MS and young binaries are in a full agreement with
the above expectations. HD155555 with $P=1.7$~d has the shortest known
period among pre-MS binaries (although some authors suggest that it may
already be on MS). All other binaries of T~Tau type and members of young
clusters with masses corresponding to CBs have periods longer than 2~d. We
conclude that theoretical, as well as observational data show that detached
binaries with initial periods of the order of 2--3~d are dominant
progenitors of CBs. Substantially shorter periods of a fraction of a day
may be encountered among young binaries only exceptionally, \eg as a result
of a hard collision with another object in a dense stellar environment but
they should be very rare (Bradstreet and Guinan 1994). Note that this
conclusion restricts an acceptable range for initial orbital periods when
modeling CBs. It is incorrect to adopt initial orbital periods shorter
than, say, 1.5~d when considering a model of a typical contact binary, as
some authors do (\eg Webbink 1976, 1977ab, Sarna and Fedorova 1989, and
more recently Jiang \etal 2012).

Binaries with components possessing subphotospheric convection zones show
hot coronae and magnetized winds carrying away mass and AM. This is a
mechanism of magnetic braking (MB). The time scale for AML of a binary with
a period of 2~d is approximately of the same order as the MS life time of
the primary component (Stêpieñ 2011a). We should expect then that a primary
is close to, or even beyond terminal age MS (TAMS) when it fills the Roche
lobe. It can fill the Roche lobe at the earlier stages of MS life only if
its mass is lower than about 1.1~\MS\ and the initial period is shorter
than 2~days (Stêpieñ and Gazeas 2012). In other words, RLOF occurs when a
typical progenitor of CB is rather old, with an age comparable to the MS
life time of its primary which under the circumstances has substantially,
or even completely depleted hydrogen in the core. Such stars lie close to
TAMS on the Hertzsprung-Russell diagram. As primaries retain their status
within the framework of the TRO model until coalescence, we should not
observe primaries of W~UMa type stars close to ZAMS. If, however, CBs are
past mass transfer with the mass ratio reversed, the present primary
consists of a little evolved former secondary and matter coming from the
outer layers of a former primary. These stars should lie close to ZAMS. The
accurate data for a number of W~UMa type stars show that most of their
primaries lie at, or close to ZAMS (Yakut and Eggleton 2005, Stêpieñ 2006a,
Siwak \etal 2010). Only the most massive primaries with masses higher than
1.5~\MS, for which the evolutionary time scale shortens substantially as a
result of increasing mass transferred slowly from secondaries, lie farther
from ZAMS. Observations of W~UMa-type stars in stellar clusters confirm
their advanced age. They are absent in young and intermediate-age clusters
while they appear abundantly in clusters with age exceeding 4--4.5~Gyr
(Rucinski 1998, 2000). Kinematical analysis of field W UMa-type stars also
shows that the binaries have an average age of several Gyr (Guinan and
Bradstreet 1988, Bilir \etal 2005).

The second basic assumption of the TRO model deals with the energy transfer
between the components. Lucy (1968, 1976) did not consider the mechanism
for it. Instead, he assumed that the transport is very efficient so that
entropies of both convection zones equalize immediately after a contact
between the components is established. Detailed mechanisms of the energy
transfer have been discussed by several authors. Applying different
simplifying assumptions, some of them argued for turbulent, small-scale
transport (\eg Moses 1976), while others considered large-scale
circulations (Hazlehurst and Meyer-Hofmeister 1973, Webbink 1977c,
Robertson 1980). However, no satisfactory result was obtained. As Yakut and
Eggleton (2005) summarized: ``... there is still remarkably little
understanding of how the heat transport manages to be as efficient as it
must be''. The problem of the energy transfer can be solved if the
assumption of the equal entropy in both convection zones is dropped (see
below).

In addition, the prediction of the TRO model that each CB should spend part
of its life as a semi-detached system finds no observational
support. Rucinski (1998) noted that CBs with distinctly different depths of
minima (suggesting poor thermal contact) are quite rare. A recent
photometric sky survey ASAS (Pojmañski 2002) detected several thousand
eclipsing binaries with periods shorter than 1~d in the solar neighborhood
(Paczyñski \etal 2006). Classification of the light curves resulted in a
significant proportion of semi-detached systems. This would seemingly solve
the problem and support TRO model, yet a closer look at the sample shows
that there is very few semi-detached binaries among stars with periods
shorter than 0.45~d (Pilecki 2010) whereas an overwhelming majority of CBs
has periods within this range (Rucinski 2007). On the other hand,
semi-detached binaries are quite common among binaries with periods between
0.7~d and 1~d where few CBs are observed. Moreover, values of global
parameters were obtained recently for several CBs and some NCBs using the
high-precision photometric and spectroscopic observations, together with
improved modeling procedure. The results show that the global parameters of
NCBs are distinctly different from those of CBs as will be demonstrated
later. So, the problem of the lack of binaries in a semi-detached phase of
TRO model still remains.

To solve the problems with the existing model of cool CBs, an alternative
model has been developed by one of us (Stêpieñ 2006ab, 2009, 2011a). The
model assumes that CBs originate from young cool binaries with initial
orbital periods close to 2~d. Both components rotate synchronously with the
orbital period and have initial masses lower than 1.3~\MS. Such stars show
strong magnetic activity resulting in MB. Permanent AML makes the orbit
tighten and the components approach each other until RLOF by the primary
occurs. Rapid mass transfer follows, up to the equalization of the
component masses. Mass transfer continues until both components regain
thermal equilibrium. The NCB configuration is identified with the mass
transfer phases when at least one component is out of thermal
equilibrium. After regaining thermal equilibrium by both components, a
contact, or Algol-type configuration is assumed. Depending on the amount of
AM left in the system, mass transfer rate and AML rate, the Algol-type
configuration may then evolve into contact or may widen the orbit and stay
as semi-detached until the presently more massive component leaves MS and a
common envelope develops. The evolution of several exemplary CBs was
computed by Stêpieñ (2006a), Gazeas and Stêpieñ (2008) and Stêpieñ and
Gazeas (2012) whereas a more systematic calculations of the cool binary
evolution from the initial state till RLOF by the primary component in a
number of systems with different initial masses and orbital periods was
carried out by Stêpieñ (2011a). Energy flow by a large scale circulation
between the components of a contact binary past rapid mass exchange was
considered in detail by Stêpieñ (2009).

The present paper investigates the relatively short-lasting but crucial
evolutionary phase of a cool close binary when the mass exchange occurs
following RLOF and the binary assumes a NCB configuration. Section~2
considers in detail the process of mass transfer including, hitherto
neglected, dynamics of the matter flowing from a donor to an accretor. The
flow forms an equatorial bulge influencing the accretion rate when the
accretor almost fills its Roche lobe. Its presence lengthens substantially
the SD1 phase of the mass transfer. This explains the observed similar
frequency of SD1 binaries, compared to SD2 binaries. In Section~3
observational data are analyzed. In particular, it is demonstrated that the
basic parameters of SD1 and SD2 binaries are very similar to each other
which favors the view that they correspond to a single process of mass
transfer. The parameters of NCBs are, on the other hand, distinctly
different from the observed parameters of CBs which contradicts an
assumption that SD1 binaries correspond to the broken contact phase of
CBs. Section~3 contains a discussion of the results and Section~4 gives
conclusions.
\vspace*{9pt}
\Section{RLOF and the Mass Transfer}
\subsection{Early Phase of Mass Transfer}
\vspace*{5pt}
Following RLOF mass transfer begins. Matter flows from a donor filling the
Roche lobe through the saddle point at the Lagrangian point L1 and falls
onto the accretor lying deep inside its Roche lobe. Most of the papers
considering the mass transfer were concentrated on the dynamics of the
stream of matter leaving L1 and its interaction with the accretor. Dynamics
of the surface layers of a donor did not attract much attention although it
does influence the mass loss process.

The problem of a surface flow in a donor was considered in an approximate
way by Lubow and Shu (1975). They demonstrated the existence of horizontal
pressure gradients with the highest pressure at the poles and the lowest at
L1. The pressure pattern results in horizontal currents similar to
geostrophic circulation driven by the Coriolis force and known from the
Earth atmosphere. The authors called this circulation
``astrostrophic''. The frictional drag exerted on the flow by lower layers
results in a slow drift of matter toward the equator. Once the matter
reaches the equator it can flow to the L1 region. This flow pattern is in
contradiction to the model considered by Webbink (1977c) who assumed that
matter leaving the donor comes from the vertical expansion of gas at the
immediate vicinity of L1. The velocity of this gas is low enough for the
Coriolis force to be negligible. Recent full 3D hydrodynamic computations
of the mass flow in the surface layers of a donor carried out by Oka \etal
(2002) contradicted the flow pattern envisaged by Webbink (1977c) and
confirmed the results of Lubow and Shu (1975). Gas elements starting at a
high astrographic latitude circle the pole where high pressure occurs,
drift to the equatorial region, run around the neck between the stars and
flow through L1. The stream leaves the donor with velocity equal to the
sound velocity and is deflected by the Coriolis force from the line joining
the centers of both stars. The same flow pattern over the donor's surface,
dominated by the Coriolis force, was obtained in numerical simulations by
Fujiwara \etal (2001).

The stream deflected by the Coriolis force misses the stellar surface and
forms a disk when the accretor radius is small enough. Otherwise the stream
strikes the surface at an inclined angle. The radial component of velocity
is dissipated and heats the in-falling matter whereas the tangential
component makes the matter flow around the star. In effect, a hot spot
appears at the impact region and an equatorial bulge encircling the
accretor is formed. The bulge is kept together by the Coriolis force. The
turbulent friction at the bottom of the equatorial stream produces the
Ekman flow to the poles so that the in-falling matter ultimately covers the
whole stellar surface whereas its excess AM (relative to the accretor) is
transferred to stellar spin.

Apart from dynamical effects, the reaction of both stars to the mass
transfer has been investigated by several authors for a range of component
masses. Mass transfer in low mass binaries, \ie possible progenitors of
W~UMa-type stars, was considered, among others, by Yungelson (1973),
Webbink (1976, 1977ab), Nakamura (1985), Sarna and Fedorova (1989),
Eggleton and Kiseleva-Eggleton (2002), Yakut and Eggleton (2005), and
recently by Ge \etal (2010). The reaction of a donor depends on its mass;
stars with masses higher than 0.9~\MS\ (assumed to be a low mass limit for
primaries in progenitors of the field W~UMa-type variables, see Stêpieñ
2006b) shrink when loosing mass (Ge \etal 2010) so, if only the mass ratio
is not too low, the mass transfer proceeds on a thermal time scale of the
donor. For the mass ratio $q=M_{\rm acc}/M_{\rm d}<1$, where $M_{\rm acc}$
and $M_{\rm d}$ are the accretor and donor mass, respectively, the accretor
reacts with swelling because its thermal time scale is longer than that of
the donor, so it can not accommodate a (too high) mass transfer rate. Note
that, due to a presence of the equatorial bulge, the accretor surface
deviates from an equipotential surface because the height of the bulge
amounts to a few percent of the stellar radius (see below). Nevertheless,
the presence of the bulge does not influence significantly the global
stellar parameters of the accretor as long as its top is within the inner
Roche lobe. The situation changes, however, when the expanding accretor
approaches the critical Roche surface.

\subsection{Mass Transfer in the Near Contact Phase}
Let us first qualitatively discuss the physical processes taking place when
the expanding accretor almost fills its inner Roche lobe.
 
The height of the equatorial bulge amounts to a few percent of the stellar
radius above the surface, so when the radius of the accretor approaches the
size of the Roche lobe by this amount, the top of the bulge begins to stick
out beyond the inner critical surface. In effect, instead of being
accreted, part of the matter flows above the Roche lobe and returns to the
donor. The accretion rate decreases until the accretor can accommodate the
accreted matter and stops the expansion. The star stays inflated with the
radius slightly smaller than the Roche lobe size and its mass increases at
a rate dictated by its own thermal time scale. If the Roche lobe descends
onto the stellar surface due to a period shortening, all matter from the
stream returns to the donor and the binary behaves as a contact system
described by Stêpieñ (2009), until the accretor shrinks and dives beneath
the Roche lobe. As a result, the accretor is surrounded by a thick
circumstellar flow of which only a small fraction adds to its mass. As time
goes on, the mass ratio approaches unity and so does the ratio of the
thermal time scales of both components. The binary temporarily assumes a
contact configuration overfilling the inner Roche lobe and, possibly, even
the outer lobe. In the latter case, a fraction of the binary mass (and AM)
may be lost from the system. In the lack of accurate models of this
process, we neglect these losses to avoid introducing arbitrary parameters
describing them. When the thermal time scale of the donor becomes longer
than that of the accretor, the latter star can accommodate the, now
decreasing, flux of matter and its radius shrinks to the thermal
equilibrium value. Also the donor returns to thermal equilibrium.
				     
We will follow Stêpieñ (2009) in a quantitative description of the flow
dynamics in the bulge close to the Roche lobe.

The fluid flow in a frame of reference rotating with a binary is described
by the continuity equation
$$\frac{\partial\rho}{\partial t}+\mathbf{\nabla\cdot}(\rho\mathbf{v})=0,\eqno(1)$$
and momentum equation
$$\frac{\partial \mathbf{v}}{\partial t}+\mathbf{(v\cdot\nabla)v}=
\frac{1}{\rho}\mathbf{\nabla}p +\nu\nabla^2\mathbf{v}
-\mathbf{\nabla}\varphi-2\Omega\times\mathbf{v}.\eqno(2)$$ 
Here $\rho, p$ and $\mathbf{v}$ denote gas density, pressure and velocity,
$\varphi$ is the gravitational (plus centrifugal) potential,$\nu$ is
kinematic viscosity coefficient and $\Omega$ is angular velocity. A full
discussion of the flow would also require the energy equation but we can
skip it when considering only dynamics of the transferred matter.

As was demonstrated by Stêpieñ (2009), the inertial term and the viscous
terms in Eq.~(2) can be neglected in contact binaries. Here, we apply the
same approximations. The resulting motion equation for a stationary flow
reads
$$\frac{1}{\rho}\mathbf{\nabla}p=2\Omega\times\mathbf{v}+\mathbf{\nabla}\varphi. \eqno(3)$$

We further ignore the region close to the neck between the stars where the
flow adjusts its size to the equilibrium condition and we concentrate on
the equatorial bulge where ${\bf v}$ has only one, non-vanishing azimuthal
component $\vv$. The equation describes the flow confined by the Coriolis
force with the width depending primarily on the flow velocity.

\begin{figure}[htb]
\centerline{\includegraphics[width=11cm]{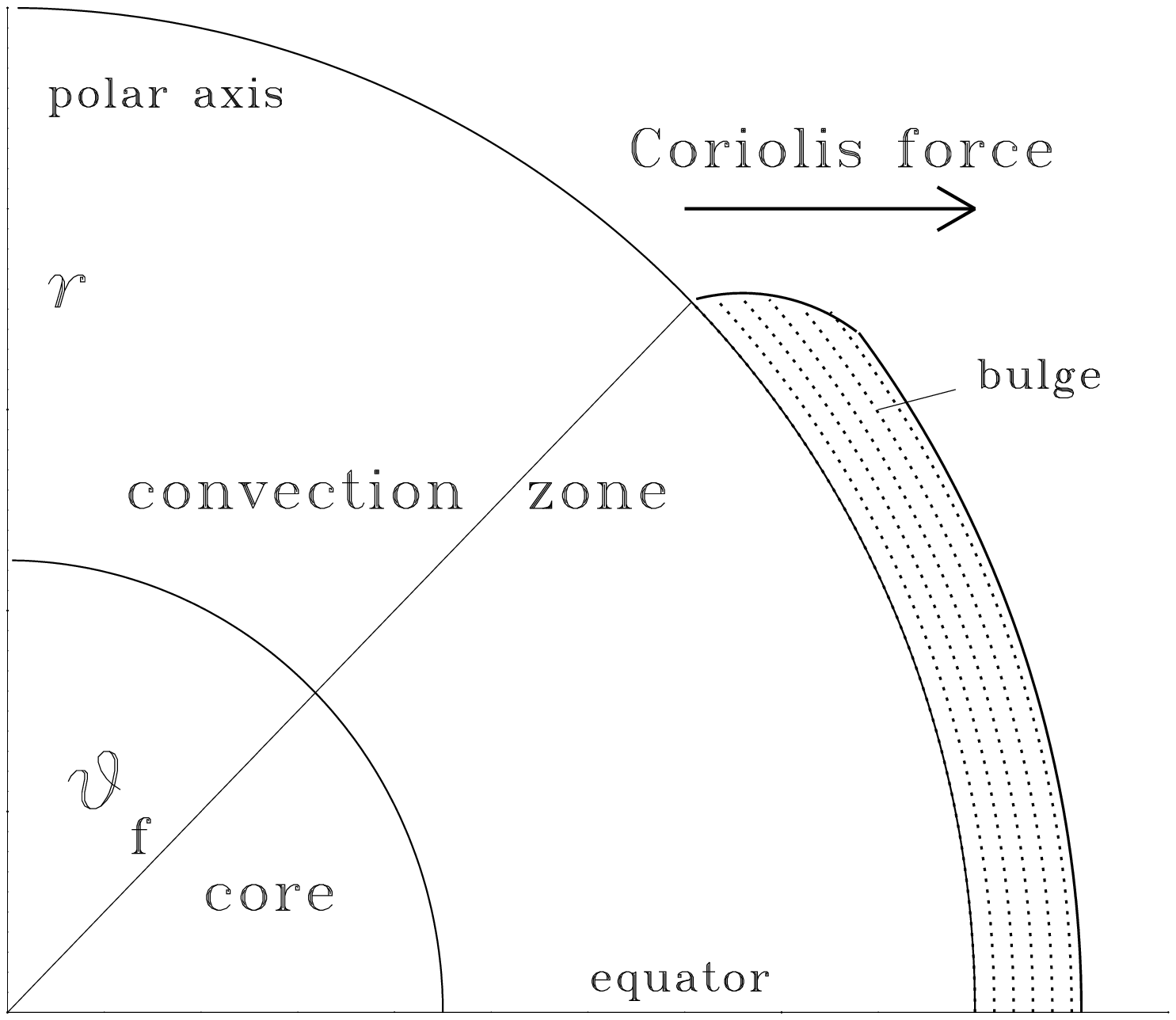}}
\FigCap{The reference frame and geometry of the flow in the meridional plane
of the accretor. The stream flows away, perpendicularly to the plane of the
paper, in the $\phi$-direction. Scale for the flow thickness has been
exaggerated.}
\end{figure}

Let us introduce a spherical system of coordinates with the origin at the
accretor center ($r,\vartheta$,$\phi$) and let us neglect the deviation
of equipotential surfaces from spheres (Fig.~1). Because the stream is
symmetric relative to the equator, we consider only a hemisphere $0\le
\vartheta\le\pi/2$. Far from the neck matter flows along the
equator of the accretor in a belt contained between equator and the
boundary given by angle $\vartheta_f(r)$ (Stêpieñ 2009)
$$\sin\vartheta_f=1-\frac{p_0}{2{\rho_0}\,\vv\,\vv_{\rm eq}}=1-\frac{c_s^2}{2\gamma\,\,\vv\,\vv_{\rm {eq}}},\eqno(4)$$
where $p_0$ and $\rho_0$ are the pressure and density inside the bulge at
the equator (they vary with the height above the accretor surface), $\vv_{\rm
eq}=\Omega R_{\rm acc}$ is the equatorial velocity of the accretor with the
radius $R_{\rm acc}$ and $c_s=(\gamma p_0/\rho_0)^{1/2}$ is the sound
velocity inside the bulge.

In an inviscid case the matter streaming from the donor is collected in the
bulge bound by the Coriolis force. As the matter accumulates, the
increasing gas pressure, density and temperature at the bottom of the bulge
results in its broadening, see Eq.~(4), until $c_s^2=2\gamma
\vv\vv_{\rm eq}$ when the bulge reaches the poles. The
non-vanishing viscosity will accelerate spreading out the bulge matter by
the Ekman drift. Molecular viscosity is negligible but because the surface
layers of the accretor, as well as the bulge, are in a turbulent state, the
turbulent viscosity will influence the flow.

The convective viscosity $\nu=l_{\rm conv}\vv_{\rm conv}/3$, where $l_{\rm
conv}$ and $\vv_{\rm conv}$ are the mixing length and convective velocity
near the bottom of the stream. The estimate of the viscosity term in Eq.~(2)
gives $\nu \vv/\Delta x^2\approx l_{\rm conv}\vv_{\rm conv}\vv/3R_{\rm
acc}^2$ so the ratio of the viscous term to the Coriolis term, $k_{\rm
drift}$, is given by
$$k_{\rm drift}=\frac{l_{\rm conv}\vv_{\rm conv}}{6R_{\rm acc}\vv_{\rm eq}}.\eqno(5)$$

Assuming that $l_{\rm conv}$ is of the order of pressure scale height near
the bottom of the bulge we obtain from the model of the convective zone
$H_p\approx 10^9$~cm and $\vv_{\rm conv}\approx2\times10^4$~cm/s
(Stêpieñ 2009). Adopting for the orbital binary period $P_{\rm
{orb}}=0.6$~d as a typical value for NCBs (see Section~3) and
$R_{\rm acc}\approx\RS$ we finally obtain $k_{\rm drift}=5.7\times10^{-6}$.
This value shows that the viscous term can indeed be neglected when
considering the azimuthal motion but the drift velocity perpendicular to
the flow motion is of the order of $k_{\rm drift}\vv$. The resulting drift
time to reach a pole is $t_{\rm drift}=R_{\rm acc}/k_{\rm
drift}\vv\approx150$~y if we adopt a value of 50~km/s for $\vv$, equal to
the sound velocity in a layer close to the inner Roche lobe of the donor
(Oka \etal 2002). The result indicates that the bulge will grow for the
first 150~y of the mass transfer and then a stationary state will be
reached with matter drifting away at the same rate as flowing in from the
donor. With a typical mass transfer rate of $1{-}2\times10^{-7}~\MS$/y
about $1{-}2\times10^{-5}~\MS$ is deposited in the bulge and its total
width is $\pm30\arcd$ from the equator, as it can be seen from
Eq.~(4). Assuming an adiabatic structure inside the bulge, its height
resulting from the model of the convection zone is equal to $\approx3\%$ of
the accretor's radius. If a back pressure along the equator is significant,
the flow velocity in the bulge may be lower than the sound speed. A
decrease of the flow velocity \eg by a factor of 2 gives a value of 300~y
for the time scale of the Ekman drift to the poles and $\pm45\arcd$ for the
width of the bulge. The change of the bulge height is insignificant.

To summarize, the presence of the bulge makes an equatorial radius of the
accretor vary between about 97\% and 100\% of the inner Roche radius during
the phase of the mass transfer before the mass ratio reversal. As long as
the bulge top is inside the Roche lobe, its temperature (besides hot spot)
should be close to the surface temperature of the accretor because the
thermal time scale of the accreted matter is much shorter than the drift
time. But when the bulge top sticks considerably above the Roche lobe and
the binary is in NCB phase, most of the matter flowing in it is replaced by
a fresh, hot matter from the donor after just one revolution around the
accretor equator. As a result, the temperature of the bulge becomes close
to the temperature of the donor surface layers (Stêpieñ 2009) whereas the
rest of the accretor surface is cooler. The mean surface temperature of the
accretor, averaged over the bulge and the polar regions, is expected to
rise relative to the unperturbed value although it still remains lower than
the donor's temperature. The deviation of the accretor shape from the
equipotential surface and the non-uniform temperature distribution make
difficult, if not impossible, unique determination of NCB geometrical
parameters, based only on the analysis of the light curve.

Accepting the TRO paradigm, some authors stopped evolutionary calculations
of the mass transfer after the accretor filled its Roche lobe and a contact
configuration was formed (\eg Sarna and Fedorova 1989, Yakut and Eggleton
2005). They assumed a rapid equalization of entropies in the convection
zones of both components resulting in thermal oscillations, with a reversed
secular mass transfer (\ie from the accretor to the donor). With an energy
transfer due to an equatorial flow, as described by Stêpieñ (2009), this
equalization does not occur and mass transfer can proceed until mass ratio
reversal, just as in case of massive binaries forming classical
Algols. Detailed models of mass transfer ignoring the assumption of the
forced equality of entropies confirm this scenario (Webbink 1976, Nakamura
1985). In particular, let us discuss details of the model calculations
carried out by Webbink (1976). He considers a binary with component masses
equal to 1.5~\MS\ and 0.75~\MS\ and the initial orbital period of
0.74~d. The evolution was strictly conservative, \ie without MB. The
primary fills its Roche lobe (and becomes a mass donor) after about
1.53~Gyr (60\% of its MS life) when the core hydrogen content has been
reduced to 0.168. Allowing for MB, binaries with initial periods of about
1.5 d would reach RLOF after a similar evolutionary advancement of the
primary (Stêpieñ 2011a) so, apart from a different evolutionary history
prior to RLOF, the results of calculations by Webbink can be applied to
some of our models.

Following RLOF, the donor transfers mass at a (maximum) rate $\dot M_d
\approx M_d/t_{\rm th}$ (Paczyñski 1971) which for the 1.5~\MS\ star
gives $\dot M_d\approx2.5\times 10^{-7}~\MS$/y. We adopt $t_{\rm
th}\approx6\times10^6$~y for a 1.5\MS\ star approaching TAMS and
$1.5\times10^8$~y for an 0.85~\MS\ star at ZAMS (0.85 results from an
initial 0.75~\MS\ mass plus 0.1~\MS\ accreted in the first phase of mass
transfer). Transferring one tenth of the solar mass takes less than one
million years. More accurate model computations give about 2~mln years
(Webbink 1976). After accreting that amount of mass the accretor fills in
its Roche lobe. Assuming that the accretor still accepts the transferred
mass, additional 0.25~\MS\ is transferred within the next 5~mln years
resulting in approximate equality of component masses, which ends the
second mass transfer phase. From now on the mass transfer rate is low
enough that the accretor can accommodate it in thermal equilibrium. It
detaches from the Roche lobe and the binary assumes an SD2
configuration. This phase takes again about 5~mln years. The next phase
of mass transfer follows with the mass transfer rate still relatively high
and the donor returning to thermal equilibrium. Because of the presently
low mass of donor, it takes about $7\times10^7$~y. Both components
reach thermal equilibrium with masses 0.96~\MS\ and 1.29~\MS\ for the donor
and accretor, respectively and with the orbital period of about 0.55~d (see
Fig.~12 and Table 2 in Webbink 1976). Mass transfer continues at a highly
reduced rate, determined only by the evolution of the donor because the
process is strictly conservative. Recent models include MB which makes the
orbit contract so that the mass transfer rate in this phase results from a
balance between the evolutionary expansion of the donor and MB. Depending
on the relative importance of these two processes, the orbital period will
shorten making the binary evolve to contact configuration or the period
will lengthen, making it evolve to the extreme mass ratio Algol. As we
argued above, the picture of the rapid mass transfer described by Webbink
(1976) is generally correct, except that the time scale of the second mass
transfer phase, prior to mass ratio reversal, is longer by an approximate
ratio of the thermal time scales of accretor and donor, \ie by about 1--1.5
orders of magnitude. This increases its duration to about $5\times 10^7$~y
in the considered case. The binary is of SD1 type at this phase.
Note that this duration is close to $7\times 10^7$~y spent in SD2
configuration (see above).

\Section{Comparison with Observations}
\subsection{Global Parameter Distributions}
According to the evolutionary scenario presented above, the NCB phase
describes the transition between detached and contact or Algol-type
binary. SD1 configuration precedes SD2 configuration and each of them takes
only a small fraction of a total cool binary life. Each binary goes only
once through both NCB phases. This is in contrast to the TRO model which
assumes that most of SD1 stars are in a broken contact phase and a binary
oscillates many times between contact and SD1 phase (Lucy 1976, Flannery
1976, Robertson and Eggleton 1977). Predictions of both evolutionary models
can be verified observationally.

Cool close binaries lose mass and AM {\it via} MB during their evolution,
so these two parameters decrease monotonically along an evolutionary
sequence. Binaries in earlier evolutionary phases are expected to have, on
average, higher values of the total mass and orbital AM, compared to
similar binaries in later evolutionary phases. Binaries in the same, or
almost the same evolutionary phase should show the same properties
regarding the orbital period, total mass and AM. In addition, the observed
frequency of stars in a given phase should be proportional to the duration
of this phase.

In Table~1 we collected available data from literature on NCBs for which
masses have been determined. Only binaries with a total mass lower than
3~\MS\ and periods shorter than 1~d were included. The 3~\MS\ limit was set
to avoid early type binaries which have different origin, do not possess
subphotospheric convection zones and are not progenitors of W~UMa-type
stars. Note that this limit is somewhat higher than 2.6~\MS\ adopted in our
earlier model calculations. Consecutive columns in Table~1 give star name,
period, component masses, orbital AM, type (SD1 or SD2) and
references. These data will be compared with the observational data on
W~UMa-type stars given by Gazeas and Stêpieñ (2008). Two contact binaries
with total masses higher than 3~\MS\ are omitted and the newer data for
AW~UMa are accepted (Pribulla and Rucinski 2008). There are 22 SD1 stars
and 27 SD2 stars in Table~1 and they will be compared with 110 W~UMa-type
stars.

There exist considerable differences in quality of both sets of data. All
W~UMa-type stars were observed spectroscopically and have good light
curves. A simultaneous solution of radial velocity and light curves results
in accurate values of stellar parameters. In case of NCB stars, very few
have radial velocity curves in addition to a light curve, resulting in
accurate values of their parameters (see \eg Siwak \etal 2010). In most
cases only a light curve of a moderate quality is available which results
in uncertain values of stellar parameters. Sometimes even for stars with
good radial velocity curves divergent data are obtained. RZ~Dra is an
example. Spectroscopic mass ratio $q=0.4$ was found by Rucinski and Lu
(2000). Yet Yang \etal (2010) found 1.91~\MS\ and 0.72~\MS\ for the
component masses ($q=0.38$) whereas Erdem \etal (2011) give 1.63~\MS\ and
0.70~\MS\ ($q=0.43$). We~ decided
\begin{landscape}
\renewcommand{\arraystretch}{0.85}
\MakeTableSepp{|c|c|c|@{\hspace{1pt}}c@{\hspace{1pt}}|c|c|
c
               |c|c|c|@{\hspace{1pt}}c@{\hspace{1pt}}|c|c|}{12.5cm}{Data on NCBs}
{
\cline{1-6}\cline{8-13}
\dorule
Name   &  Period [d] & Masses [\MS] & $H_{\rm orb}{\cdot}10^{51}$ [cgs] & Type &  Lit. &&
\uprule
Name   &  Period [d] & Masses [\MS] & $H_{\rm orb}{\cdot}10^{51}$ [cgs] & Type &  Lit. \\
\cline{1-6}\cline{8-13}
\uprule
BL And    & 0.7224 & 1.80+0.70 & 10.316 & SD1  & Z06  && DZ Cas    & 0.7849 & 1.89+0.17 & 2.882  & SD2  & Y12b \\
CN And    & 0.4628 & 1.43+0.55 & 5.995  & SD1? & S10  && IV Cas    & 0.9985 & 1.98+0.81 & 14.077 & SD2  & K10  \\
NP Aqr    & 0.8070 & 1.65+0.99 & 13.660 & SD1  & I10  && EG Cep    & 0.5446 & 1.65+0.77 & 9.554  & SD2  & Z09a \\
V609 Aql  & 0.7966 & 1.05+0.72 & 7.193  & SD1  & T08  && YY Cet    & 0.7905 & 1.78+0.92 & 13.479 & SD2  & W12  \\
DO Cas    & 0.6847 & 2.13+0.65 & 10.735 & SD1  & S10  && TW CrB    & 0.5889 & 1.20+0.87 & 8.518  & SD2? & F12  \\
V473 Cas  & 0.4155 & 1.00+0.48 & 3.911  & SD1? & Z09b && W Crv     & 0.3881 & 1.00+0.68 & 5.182  & SD2  & R00  \\
V747 Cen  & 0.5371 & 1.54+0.49 & 6.017  & SD1? & S10  && RV Crv    & 0.7473 & 1.64+0.45 & 6.473  & SD2  & M86a \\
VV Cet    & 0.5224 & 1.04+0.30 & 2.822  & SD1? & S10  && V836 Cyg  & 0.6534 & 2.20+0.78 & 12.807 & SD2  & Y05  \\
WZ Cyg    & 0.5845 & 1.59+1.00 & 11.971 & SD1? & S10  && RZ Dra    & 0.5509 & 1.77+0.71 & 9.226  & SD2  & E11  \\
BV Eri    & 0.5077 & 1.56+0.43 & 5.283  & SD1? & S10  && AX Dra    & 0.5681 & 1.46+0.92 & 10.342 & SD2? & K04  \\
TT Her    & 0.9121 & 1.56+0.68 & 9.741  & SD1  & M89  && FG Gem    & 0.8191 & 1.75+0.71 & 10.681 & SD2  & LN11 \\
FS Lup    & 0.3814 & 1.30+0.61 & 5.740  & SD1? & S10  && GW Gem    & 0.6594 & 1.74+0.80 & 11.013 & SD2  & L09  \\
FT Lup    & 0.4701 & 1.87+0.82 & 10.628 & SD1? & S10  && SZ Her    & 0.8181 & 1.58+0.75 & 10.373 & SD2  & L12  \\
SW Lyn    & 0.6441 & 1.72+0.90 & 11.998 & Det? & KR03 && AV Hya    & 0.6829 & 2.38+0.62 & 11.155 & SD2  & Y12a \\
UU Lyn    & 0.4685 & 2.10+0.74 & 10.578 & SD1? & Z07  && DD Mon    & 0.5680 & 1.29+0.87 & 8.925  & SD2  & P09  \\
V361 Lyr  & 0.3096 & 1.26+0.87 & 7.149  & SD1  & H97  && DI Peg    & 0.7118 & 1.18+0.70 & 7.402  & SD2  & LU92 \\
V1010 Oph & 0.6614 & 1.89+0.89 & 12.914 & SD1  & S10  && HW Per    & 0.6348 & 1.66+1.10 & 13.836 & SD2  & S98  \\
BO Peg    & 0.5804 & 1.90+1.00 & 13.776 & SD1  & Y86  && AG Phe    & 0.7553 & 1.42+0.22 & 2.997  & SD2  & C96  \\
RT Scl    & 0.5116 & 1.63+0.71 & 8.635  & SD1  & H86  && ROSAT (1) & 0.3881 & 0.92+0.82 & 5.633  & SD2  & LI09 \\
GR Tau    & 0.4299 & 1.45+0.32 & 3.590  & SD1  & G04  && V1241 Tau & 0.8230 & 1.60+0.87 & 11.950 & SD2  & Y12b \\
V1374 Tau & 0.2508 & 0.67+0.32 & 1.680  & SD1? & A11  && BF Vel    & 0.7040 & 1.98+0.84 & 12.960 & SD2  & M09  \\
BS Vul    & 0.4760 & 1.52+0.52 & 6.050  & SD1  & Z12  && BF Vir    & 0.6406 & 1.23+0.41 & 4.569  & SD2  & R81  \\
CX Aqr    & 0.5560 & 1.19+0.64 & 6.363  & SD2  & M86b && FO Vir    & 0.7760 & 2.00+0.26 & 4.522  & SD2? & R05  \\
ZZ Aur    & 0.6012 & 1.62+0.76 & 9.643  & SD2  & O06  && AW Vul    & 0.8065 & 1.60+0.87 & 11.901 & SD2  & LN11 \\
HL Aur    & 0.6225 & 1.50+1.10 & 12.686 & SD2? & Z97  &&           &        &           &        &      &      \\
\cline{1-6}\cline{8-13}
\noalign{\vskip3pt}
\multicolumn{13}{p{20cm}}{ROSAT(1)=1RXS~J201607.0251645
Lit.:
A11=Austin \etal 2011,
C96=Cerruti 1996,
E11=Erdem \etal 2011,
F12=Faillace \etal 2012,
G04=Gu \etal 2004,
H86=Hilditch and King 1986,
H97=Hilditch \etal 1997,
I10=Ibanoglu \etal 2010,
K04=Kim \etal 2004,
K10=Kim \etal 2010,
KR03=Kreiner \etal 2003,
L09=Lee \etal 2009,
L12=Lee \etal 2012,
LI09=Li \etal 2009,
LN11=Liakos \etal 2011,
LU92=Lu 1992,
M09=Manimanis \etal 2009,
M86a=McFarlane \etal 1986a,
M86b=McFarlane \etal 1986b,
M89=Milano \etal 1989,
O06=Oh \etal 2006,
P09=Pribulla \etal 2009,
R00=Rucinski and Lu  2000,
R05=Rucinski \etal 2005,
R81=Russo and Sollazzo 1981,
S10=Siwak \etal 2010,
S98=Samec \etal 1998,
T08=Turner \etal 2008,
W12=Williamon and Sowell 2012,
Y05=Yakut \etal 2005,
Y12a=Yang \etal 2012a
Y12b=Yang \etal 2012b
Y86=Yamasaki and Okazaki 1986,
Z06=Zhu and Qian 2006,
Z07=Zhu \etal 2007,
Z09a=Zhu \etal 2009a,
Z09b=Zhu \etal 2009b,
Z12=Zhu \etal 2012,
Z97=Zhang \etal 1997
}}
\end{landscape}
\noindent
to list arithmetic means of both results in Table~1. It is well known that the
mass ratio is poorly constrained when only light curve is available,
particularly for low orbit inclinations. In addition, the commonly used
Wilson-Devinney program for solving light curves gives often ambiguous
solutions with marginally different quality of fit between SD1 and SD2
configuration (Qian \etal 2006) or even among NCB, contact and detached
configuration (Odell \etal 2009). The ambiguity results in multiple
classification of some stars, depending on subtle details of the analyzed
light curve or, possibly, on the intention of the author(s) who may \eg
prefer the SD1 solution for a binary with a decreasing period. Examples of
divergent classifications are given by Siwak
\etal (2010). All three data samples suffer from an observational bias
toward bright objects. Nevertheless, we believe that the samples are
sufficiently representative to draw meaningful conclusions about their
statistical properties.

We first compare the observed frequencies of binaries in different
evolutionary phases with the expectations resulting from our model. The
analysis of V361~Lyr by Hilditch \etal (1997) indicates that the binary is
in the initial phase of mass transfer when the accretor has not yet
expanded in reaction to the matter flowing onto it. The light curve shows a
prominent hump caused by a hot spot on the accretor surface. The star
should approach its Roche lobe in about 2~mln years as the calculations by
Sarna and Fedorova (1989) show. Similar time scale was obtained by Webbink
(1976) and Nakamura (1985). Recently, another faint star, 2MASS
J05280799+7256056, was detected showing a light curve with a similar hump
(Virnina \etal 2011). According to the ``classical'' model by Webbink
(1976) and Nakamura (1985) the next phase of the mass transfer, until the
component mass equalization, takes only 2--3 times longer than the initial
phase. We suggest that it takes about 1--1.5 orders of magnitude
longer. The observed number of the V361~Lyr-type variables (one, or
possibly two) can be compared to the number of the other SD1 variables
listed in Table~1, which is 21. Their ratio is not in contradiction with
our predictions and suggests a much longer duration of the total SD1 phase
than that of V361~Lyr-like.

A similar comparison can be made for the duration of SD1 \vs SD2 phase. Our
scenario predicts that each of them takes approximately the same time
interval. This is in a fair agreement with the numbers of binaries in each
configuration, listed in Table~1: 22 of SD1 \vs 27 of SD2 binaries.

Our model assumes that SD2 phase follows SD1 phase and that they both last
rather shortly, compared to the total evolutionary life time of a
binary. For a conservative mass exchange, global parameters of a binary
should not change significantly between these two phases. We can check this
by comparing both samples of NCBs. The average total mass of SD1 binaries
listed in Table~1 is equal to $2.20\pm0.11$ and of SD2 binaries to
$2.32\pm0.08$. Both numbers differ insignificantly. The same conclusion is
drawn regarding the mean values of the orbital angular momentum. Here we
have $8.20\pm0.76$ and $9.23\pm0.64$ ($\times10^{51}$ in cgs units), for
SD1 and SD2 binaries, respectively. Only the average orbital periods of
both groups of NCBs differ significantly: $0.55\pm0.03$~d \vs
$0.67\pm0.03$~d, for SD1 and SD2, respectively. However, orbital period is
not a monotonic function of time. As we explained above, some of SD1
binaries evolve quickly to a contact configuration by shortening their
periods (or, possibly, by mass and AML in a non-conservative case of mass
transfer) whereas the others evolve to an Algol configuration by
lengthening their periods (Stêpieñ 2011b). The former binaries fall out
because they are classified as CBs. On the other hand, the boundary between
stars classified as SD2-type binaries and short-period Algols is not well
defined. As a result, there are SD2 binaries listed in Table~1 with a
primary filling only about 80\% of its Roche lobe (SZ~Her or V1241~Tau) and
longer than average periods. Had we moved them to a category of usual
Algols, we would have obtained a lower value for an average period of SD2
binaries. So, we do not consider the difference in average periods of SD1
and SD2 binaries as meaningful.

\begin{figure}[htb]
\vglue-3mm
\centerline{\includegraphics[width=11cm]{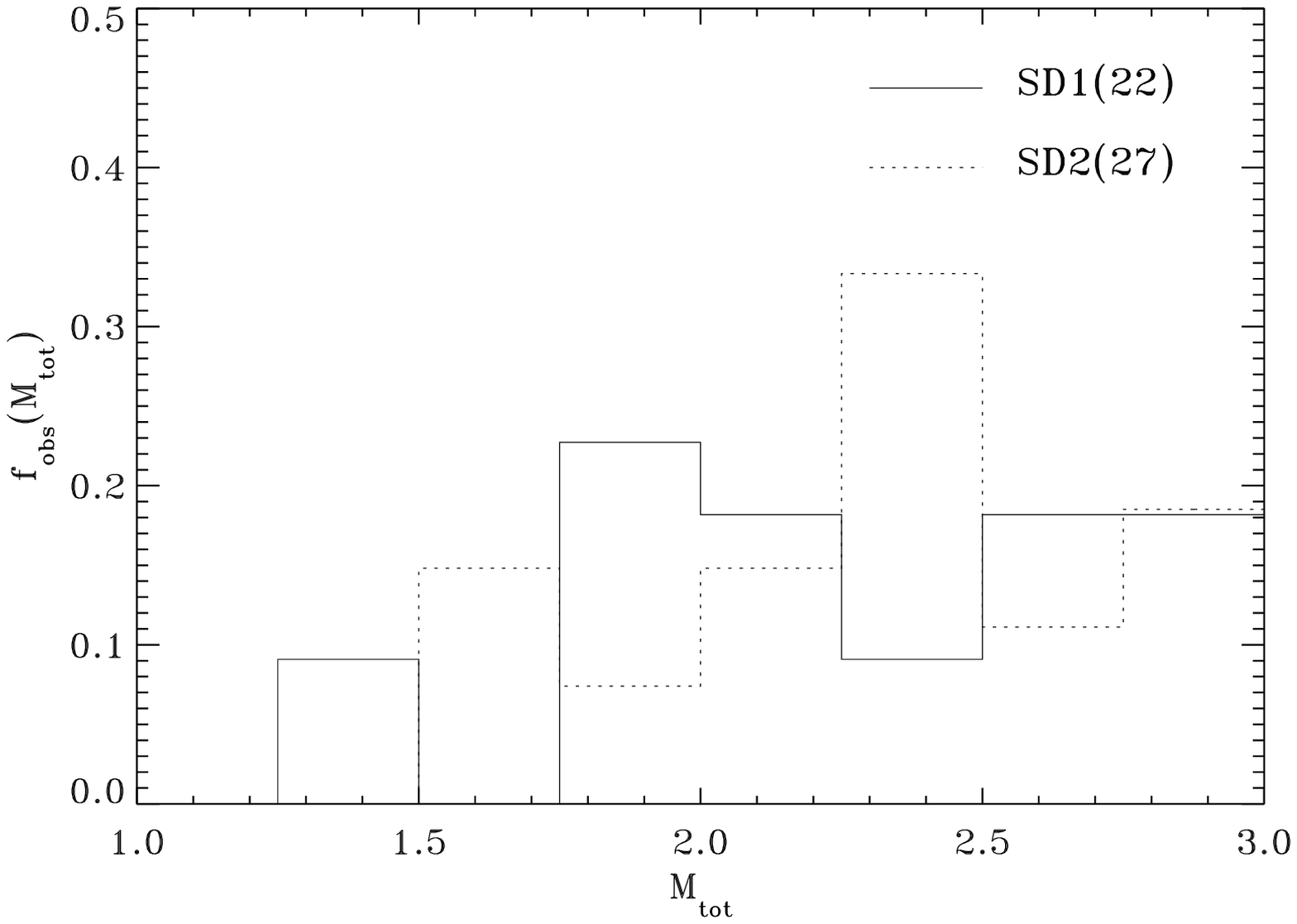}}
\vspace*{-3mm}
\FigCap{Mass distributions of NCBs from Table~1. Both distributions are 
normalized to unity. Solid and dotted lines describe data for SD1 and SD2 
binaries, respectively.}
\end{figure}
\begin{figure}[htb]
\centerline{\includegraphics[width=11cm]{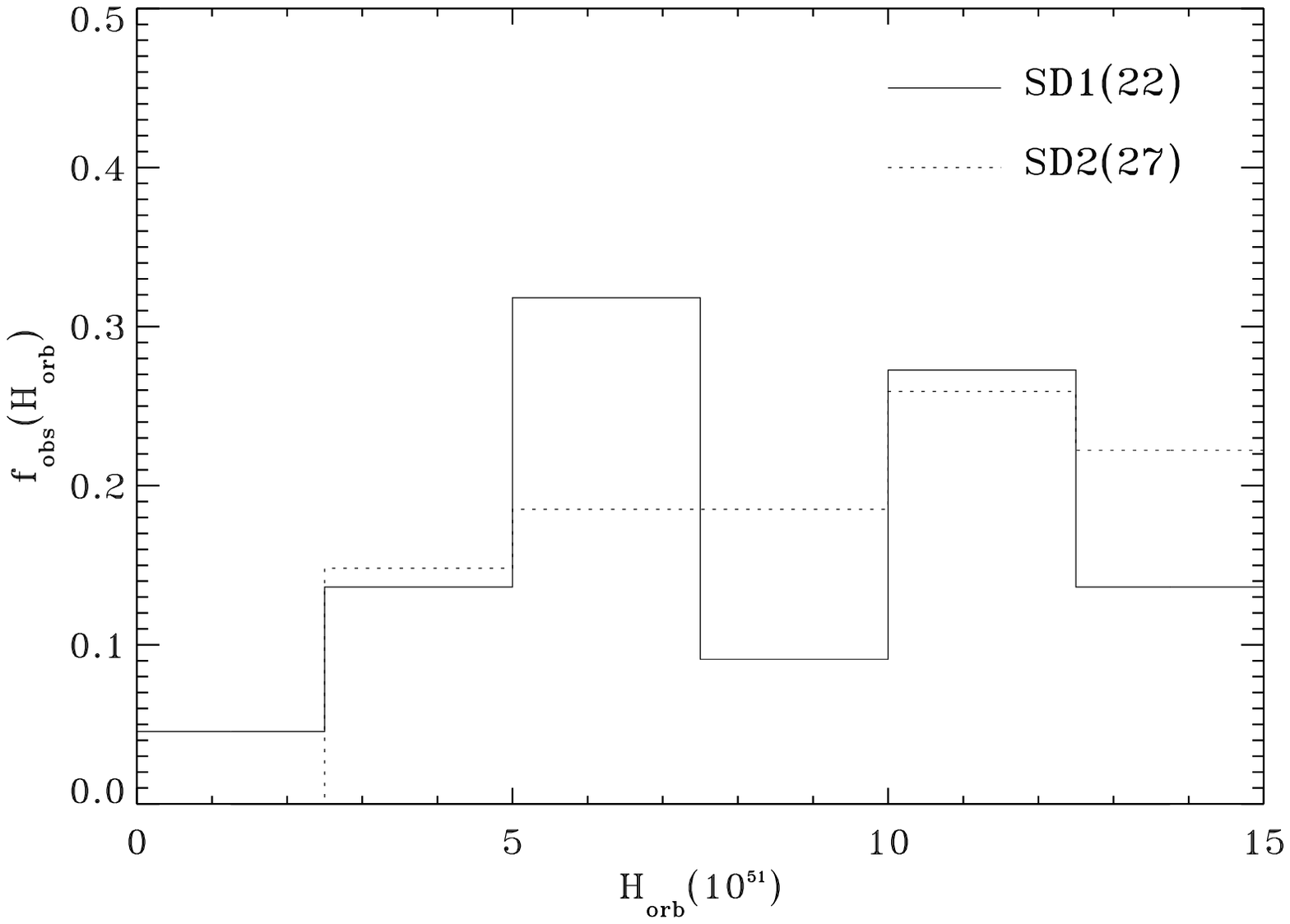}}
\vspace*{-3mm}
\FigCap{Orbital angular momentum distributions of NCBs from Table~1.}
\end{figure}
\begin{figure}[htb]
\centerline{\includegraphics[width=11cm]{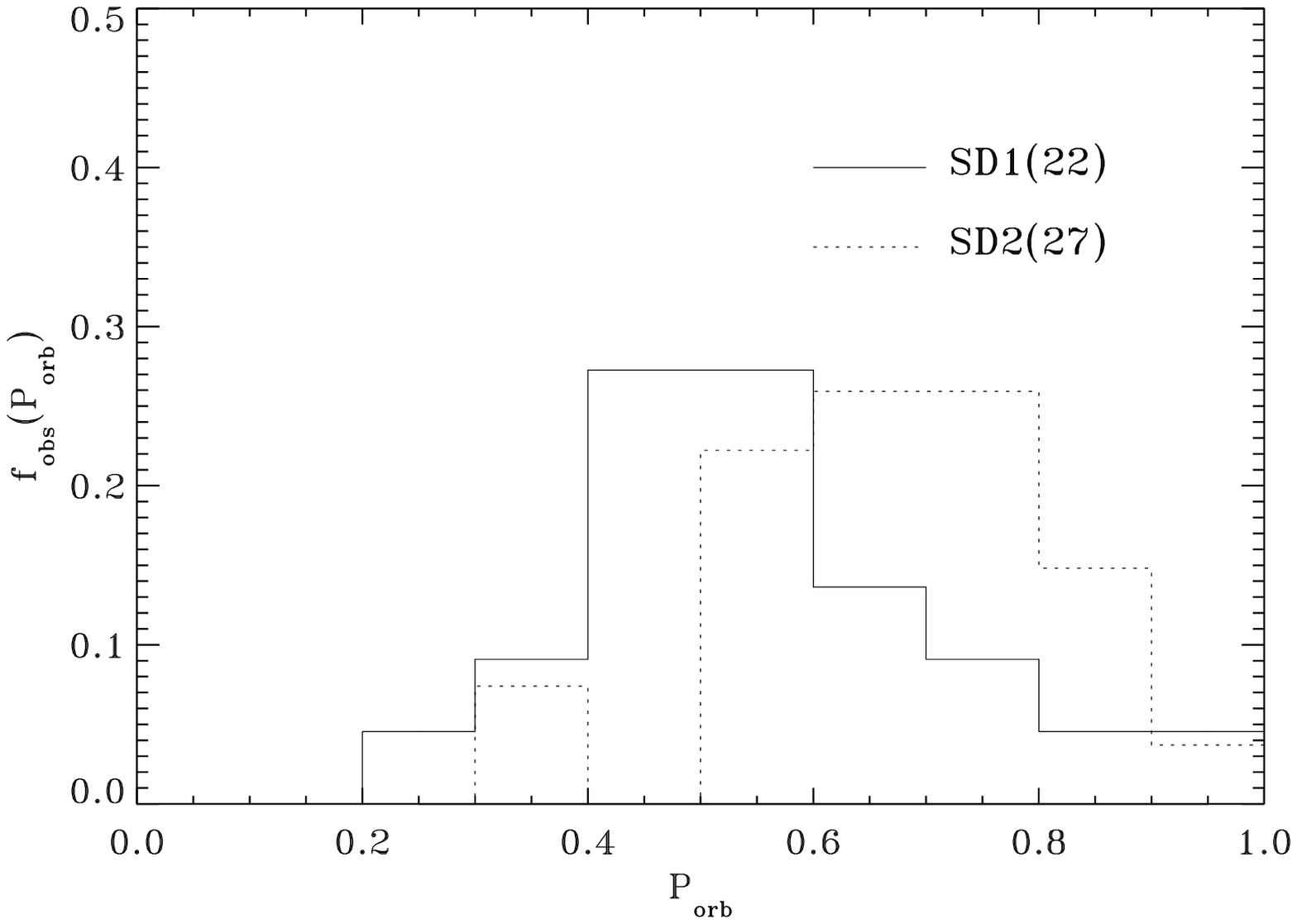}}
\vspace*{-3mm}
\FigCap{Orbital period distributions of NCBs from Table~1.}
\end{figure}
With sufficiently numerous samples it is possible to compare not only mean
values but also the distributions of the discussed parameters. Figs.~2, 3
and 4 show the mass, AM and period distributions of both kinds of NCBs. It
is easy to see that the respective distributions look very similar. The
statistical analysis supports this impression. The $\chi^2$ -- test applied
to all three pairs of the distributions gives values of the reduced
$\chi^2$ equal to 1.71, 0.64 and 1.65 for mass, AM and period
distributions, respectively. Null hypothesis that both samples were drawn
from different distributions can be rejected at the 5\% level, although the
probability of different distributions in case of the total mass and period
is slightly higher than in case of AM.

\begin{figure}[htb]
\centerline{\includegraphics[width=11cm]{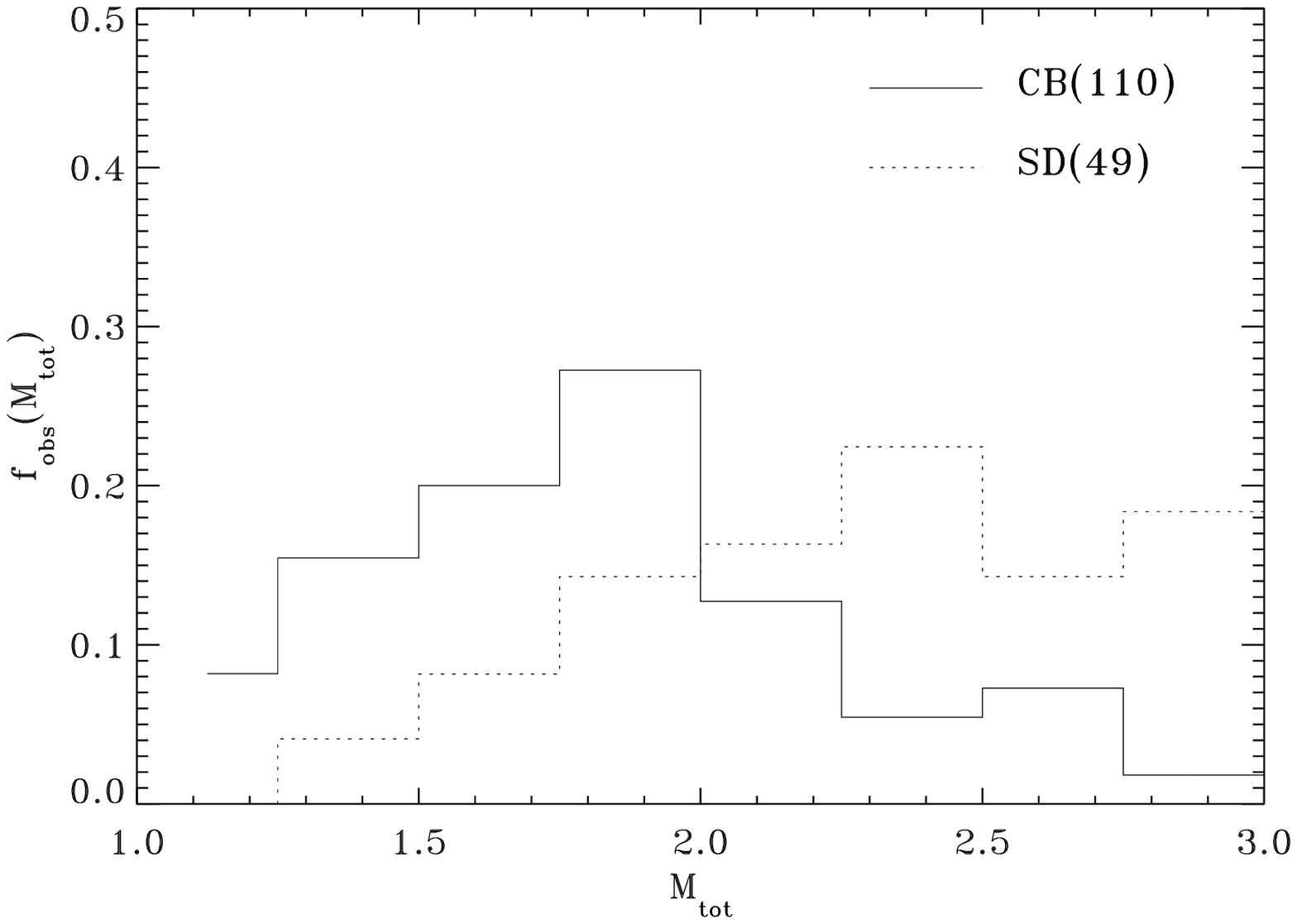}}
\vspace*{-3mm}
\FigCap{Mass distributions of all NCBs from Table~1 and CBs from the paper
by Gazeas and Stêpieñ (2008), see text. Both distributions are normalized
to unity. Solid and dotted lines describe data for CBs and NCBs,
respectively.}
\end{figure}
\begin{figure}[htb]
\centerline{\includegraphics[width=11cm]{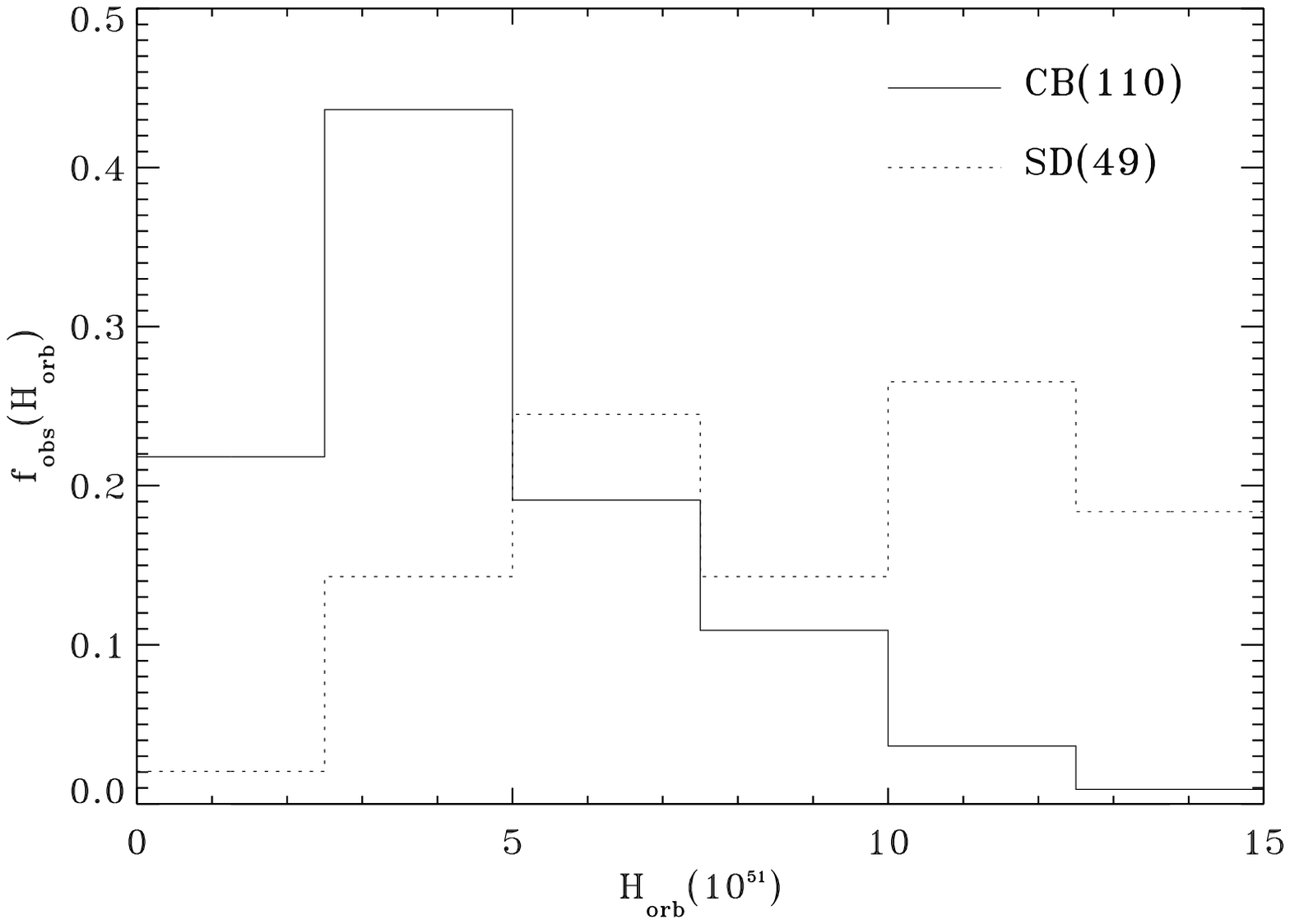}}
\vspace*{-3mm}
\FigCap{Orbital angular momentum distributions of all NCBs from Table~1 and
CBs from the paper by Gazeas and Stêpieñ (2008).}
\end{figure}
\begin{figure}[htb]
\centerline{\includegraphics[width=11cm]{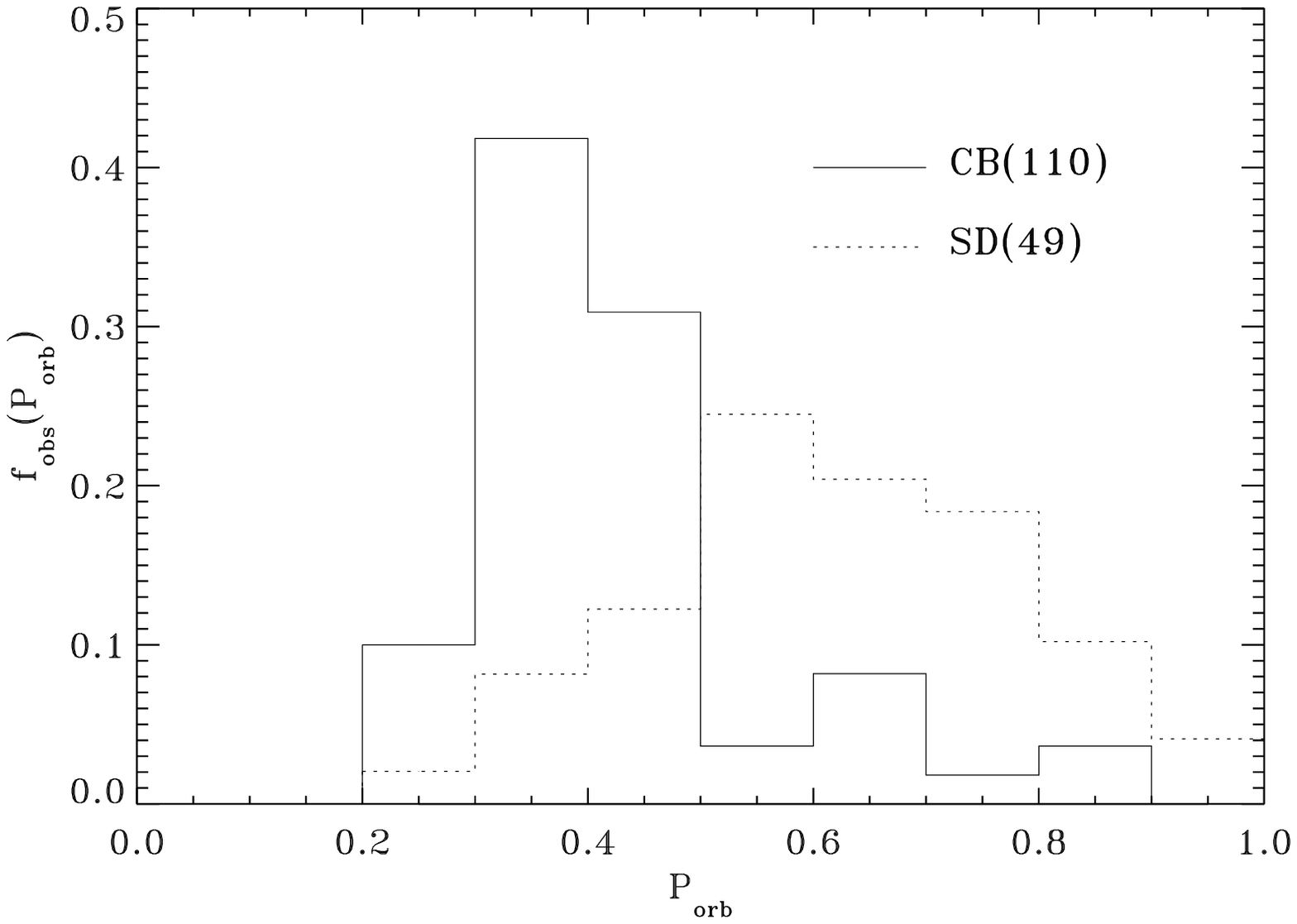}}
\vspace*{-3mm}
\FigCap{Orbital period distributions of all NCBs from Table~1 and CBs from
the paper by Gazeas and Stêpieñ (2008).}
\end{figure}

We conclude that no significant differences exist between both samples of
NCBs. So, we merge them together and compare with CBs from Gazeas and
Stêpieñ (2008), as described above. A mean value of the total mass of CBs
is equal to $1.81\pm0.04$~\MS\ which is substantially less than
$2.27\pm0.06$~\MS\ -- a mean value for all NCBs. A mean value of the orbital
AM of CBs is equal to $4.68\pm0.25$ which can be compared to 8.$73\pm0.50$
-- a mean value for NCBs. Again, the difference is substantial. Finally,
mean values of the period are $0.42\pm0.01$~d and $0.61\pm0.02$~d, for CBs
and NCBs, respectively. They also are substantially different. The TRO
model predicts period variations in the course of an oscillation but the
expected amplitude does not exceed 10\% (K\"ahler 2002). The observed
difference is several times larger. Figs.~5, 6 and 7 show the distributions
of the respective parameters for both groups of binaries. It is immediately
visible that the distributions are different. The reduced $\chi^2$ values
are equal to 7.12, 11.21 and 9.43 for total mass, AM and period,
respectively. So high values of $\chi^2$ mean a negligible probability that
both samples were drawn from the same distribution.

We conclude that NCBs are in a different (presumably earlier) evolutionary
stage than CBs. This leaves no candidates for a broken contact binaries
predicted by TRO model but is in agreement with our model where CBs are
individually in thermal equilibrium and stay in contact until component
merging.

\subsection{Period Variations}
According to the TRO theory length of the orbital period of a CB varies in
the course of a thermal oscillation. With a typical period of 0.3~d and a
thermal time scale of $10^6{-}10^7$~y we expect the time derivative of the
period, $\dd P/\dd t$, to have typical values between a few times $10^{-8}$
up to a few times $10^{-7}$, in units of d/y. Period variations of the same
order are expected in the broken contact binary during the oscillation.
Indeed, period variations of this order have been reported for several CBs
and NCBs (Qian 2001ab, Zhu and Qian 2006, Zhu \etal 2009a, see also
references to individual objects in Table~1). Can this be a proof for
correctness of the TRO theory? Hardly, in our opinion. Here are some
counterarguments.

Most of the individual values of period variations was determined from
analysis of the $O-C$ diagrams. A second order polynomial is fit to the
observational data and a period variation is described by the quadratic
term. Errors of the quadratic coefficient are seldom given by the author
analyzing the $O-C$ diagram but if the error is given, it is typically just
a few times smaller than the coefficient itself. There are, however,
several cases when no apparent trend is visible in the $O-C$ diagram
(see \eg Kreiner \etal 2001). These cases have usually been ignored when
discussing period variations of a sample of variables. Obviously, such
samples are strongly biased toward binaries with large period
variations. Only recently, two systematic surveys of period variations of
CBs have been carried out; Kubiak \etal (2006) analyzed 569 CBs observed by
the OGLE team and Pilecki (2010) analyzed almost 6000 short-period
eclipsing binaries (among them CBs, NCBs and detached binaries) observed
within the ASAS program. Both analyzes showed that only about one third of
the objects show measurable period variations. For the rest of stars any
variations are below the accuracy of determination. Unfortunately, the
uncertainties of both surveys were still quite considerable due to a short
duration of OGLE and ASAS programs. Nonetheless, any statement about the
presence and magnitude of secular period variations in CBs and NCBs seems
premature and must await more systematic investigation.

In addition to any possible systematic trend, $O-C$ diagrams of many
binaries show a wavy behavior interpreted as periodic, or quasi-periodic
orbital period variation. Two possible mechanisms are considered in respect
to them: presence of a third body or variation of stellar moment(s) of
inertia due to magnetic activity cycles, as suggested by Applegate (1992).
Third body is usually preferred (Liao and Qian 2010) because almost all
cool short-period binaries are expected to have such a companion (see
Introduction). The resulting periods of outer orbits range from a few years
up to the length of the time interval covered by $O-C$ diagrams. Liao and
Qian (2010) even claim a detection of a cyclic period variation of WW~Dra
with the length of 112~y. Because of the ubiquitous presence of third
companions, interpretation of $O-C$ diagrams of CBs and NCBs needs the
utmost care as stressed by Hilditch (1989). Many apparently parabolic
diagrams show a sudden turn when additional observations are added and
instead of a secular period variation a cyclic variation appears
(J. Kreiner -- private communication). Lack of any secular period variation
is reported for the following objects from Table~1: NP~Aqr, DO~Cas,
V747~Cen, VV~Cet, FS~Lup, SW~Lyn, V1374~Tau of SD1 type and IV~Cas, YY~Cet,
AX~Dra, FG~Gem, SZ~Her, DI~Peg, HW~Per, V1241~Tau and AW~Vul of SD2
type. In most these cases a cyclic variation is present which is
interpreted in terms of the tertiary system, although some authors also
consider the Applegate mechanism (Zavala \etal 2002).

There exists yet another possible mechanism of period variations of CBs and
NCBs. The energy transport between the components of a CB or SD1 binary
requires about $10^{-4}{-}10^{-5}$~\MS/y of matter flowing back and forth
between the components (Webbink 1977c, Martin and Davey 1995, Stêpieñ
2009). Random fluctuations of this flow, resulting \eg from the variable
magnetic activity, at the relative level of $10^{-3}$, will result in
period variations of the order of $10^{-7}$~d/y. Weaker fluctuations
will produce, of course, correspondingly weak period variations. All these
variations will have a random distribution. Kubiak \etal (2006) concluded
that the observed distribution of period variations of CBs can, indeed, be
described by the Gaussian curve. This indicates a purely random mechanism
for these variations.

We conclude that the observed period variations of CBs and NCBs can be
caused by so many different mechanisms that their presence in individual
binaries cannot be used as an evidence for the existence of TRO.

\Section{Discussion}
The problem of evolutionary connection between W UMa-type variables and
other cool short-period binaries has been discussed since their
discovery. Hilditch (1989) summarized his earlier investigations of several
contact and semi-detached stars with the conclusion that there must exist
two ways of forming cool CBs: first, when a moderate mass transfer from a
primary makes a secondary also fill up quickly its Roche lobe, forming in
effect a contact binary of W-type in a state of TRO, and second, when the
mass transfer proceeds until mass ratio reversal so that an Algol-type
configuration is first formed, which next transforms into a
marginal-contact system and then into a deep-contact A-type system. Note,
that the Kuiper paradox does not apply to the latter binaries, as was
already mentioned in Section~1, so each component can individually be in
thermal equilibrium and fit to its Roche lobe. There is no need for
artificially inflated secondary (former primary) by elevated value of its
specific entropy in this case, although energy must be transported to it
from the present primary. If Hilditch was right, there would exist a
fundamental difference in the physical properties of A-type and W-type
variables. However, as we know now, the division between two types has no
fundamental physical meaning and is most probably connected with the degree
of spottiness of the primary component. As Gazeas and Niarchos (2006)
demonstrated, the most massive CBs with periods longer than 0.5~d are only
of A-type. Their primaries show very little or no magnetic activity and
they are permanently hotter than their companions. The least massive CBs
with periods shorter than 0.3~d are, on the other hand, only of
W-type. Here, the primaries possess deep convection zones and are expected
to be heavily spotted which decreases their mean surface temperatures. The
secondaries are very likely much less spotted (Stêpieñ \etal 2001). Both,
A- and W-types occur for CBs with the intermediate masses and periods
between 0.3~d and 0.5~d. Differences between the mean surface temperatures of
both components (they decide about the classification of the variables as
an A-type or W-type) are rather small among these binaries and change
sometimes sign, which makes some binaries to alternate between the types.

The idea of an evolutionary sequence of cool CBs: detached
${\rm \rightarrow SD1\rightarrow SD2 \rightarrow}$ 
CB was later
suggested by Shaw (1994). Accurate stellar parameters were then available
for very few NCBs so the author based his proposition on the apparent
similarity of SD2 type stars to the A-type CB. He noted that a moderate
shortening of the period would bring an SD2 binary into contact and with
the efficient energy transport between the components it would simply
become an A-type CB. He also noted that a typical light curve of an SD1
variable shows asymmetry resulting presumably from a hot spot on the
trailing side of the cool component due to the mass transfer through the
inner Lagrangian point from the hot component. He concluded that SD1 stars
are nearer the beginning of their evolution toward CBs and after mass
ratio inversion they become SD2 stars. We confirm here his proposition
quantitatively with the accurately known stellar parameters of sufficiently
numerous samples of cool close binary stars, as shown in Section~3.

The problem of an evolutionary status of NCBs was also considered by Yakut
and Eggleton (2005). Based on available at that time observations they
determined stellar parameters of several cool close binaries of different
types, compared them with each other and with evolutionary models obtained
earlier by Eggleton and collaborators. In particular, they noticed that the
total mass and angular momentum of NCBs are significantly higher than of
CBs. However, they tried to explain this discrepancy within the TRO model
by making an arbitrary assumption that massive CBs with long periods spend
more time in the broken contact phase (identified with the SD1
configuration) as opposite to low mass CBs which must spend very little (if
any) time in the broken contact phase because so few SD1 binaries is
observed with short periods. Yakut and Eggleton (2005) also discuss
evolutionary status of SD2 binaries. Following Hilditch (1989) they suggest
that these are binaries past case A mass transfer, \ie with a primary
filling the Roche lobe when it is still on MS. In other words, SD2 binaries
are presently in the Algol configuration. But what about early phases of
mass transfer in these binaries, before mass ratio reversal?  They argue
that some of the observed SD1 binaries are in fact
``first-timers''. Because, however, this phase takes a much shorter time
according to a standard mass transfer model (Webbink 1976, Nakamura 1985,
see also above) we should observe several times more SD2 than SD1
binaries. The observations show otherwise as we know and as also Yakut and
Eggleton noted; both groups of stars are roughly equally numerous. The
authors concluded that the majority of SD1 binaries must be in a broken
contact phase of TRO. Unfortunately, they could not distinguish between
these two groups of SD1 binaries and were unable to offer any observational
test to do so. This apparent ambiguity in evolutionary status of SD1
binaries leads authors analyzing individual objects of this type to the
standard conclusion in a form of a rhetoric question: is the analyzed
variable a first-timer or a CB in a broken contact phase (see references to
Table~1). With our model of mass transfer described in Section~3, all SD1
binaries are first-timers and they should be as numerous as SD2 stars. As
mentioned earlier, a short phase between SD1 and SD2 configuration exists
when both components have almost equal masses and are in contact. FT~UMa
may be an example of such CB with a total mass less than 3~\MS\ (Yuan
2011). The binary has a period of 0.65~d, component masses of 1.49~\MS\ and
1.46~\MS\ ($q= 0.98$), and radii of 1.79~\RS\ and 1.78~\RS. The existing
observational data do not allow to decide whether the binary is still prior
to, or just past mass reversal.

New data on CBs and NCBs obtained in the recent years, based on systematic
and accurate photometric and spectroscopic surveys of cool close binaries
like ASAS, OGLE and DDO, did not resolve problems inherent in the TRO
model. On the contrary, the problems seem to be even more severe than
ever. Adoption of the idea that all cool CBs are past mass ratio
reversal, \ie they are in a similar evolutionary state as short-period
normal Algols, explains all important observational facts in a consistent
way. CBs form a low-mass, short-period end of a sequence of binaries
showing so called Algol paradox (when the less massive component is
evolutionary more advanced than its more massive companion). Both
components of such configuration are in thermal equilibrium. In addition, a
common envelope surrounding both components of a CB enforces circulation
carrying energy from the more to less massive component, which results in
an apparent equality of the average surface brightness.

The model presented in this paper can be verified by direct hydrodynamical
simulations of the mass transfer in CBs and NCBs. Two-dimensional
computations of a mass transfer in a CB carried out by Martin and Davey
(1995) demonstrated the existence of an asymmetric stream deflected by the
Coriolis force and encircling the secondary component, but two dimensions
restricted severely dynamics of the flows in the common envelope. Full 3D
calculations are necessary, as shown by Oka \etal (2002). It is apparent
from their results that dynamics of matter in surface layers of a donor is
dominated by the Coriolis force just as is dynamics of air in the Earth
atmosphere. Including an accretor, first under-filling and then overfilling
its critical Roche lobe, into the computational domain would greatly
contribute to our understanding of the mass and energy exchange between the
components of NCBs and CBs.

With fast developing techniques for measuring velocity fields over the
stellar surface we may soon be able to obtain detailed maps of mass motions
in NCBs and CBs. Approximate information can already be obtained from the
analysis of the spectral line profiles as was done \eg by Rucinski and his
collaborators. As an example, the broadening function of AW~UMa indicates
that the radial velocity of the polar regions in both components is lower
than, but in the equatorial regions is close to velocity expected for stars
filling their Roche lobes and rotating synchronously with the orbital
period (Pribulla and Rucinski 2008). This peculiar velocity field can be
explained by our model (Stêpieñ 2009). Another model prediction concerns
the temperature distribution over the surface of a low mass component in
CBs and SD1 binaries with a hotter equatorial belt and cooler polar
caps. The accurate temperature mapping should be able to verify this
prediction.

Finally, we stress that if the future observations confirm the existence of
the equatorial bulge in CBs and SD1 binaries, their conventional model
assuming strict hydrostatic equilibrium and stellar surfaces lying exactly
on equipotential surfaces must be abandoned. Dynamical effects will have to
be included when modeling these stars.

\Section{Conclusions}
The observed properties of cool close binaries with one component filling,
and the other nearly filling its Roche lobe (called NCBs) fit to the recent
model of origin and evolutionary status of cool CBs developed by one of us
(Stêpieñ 2006ab, 2009, 2011a, Gazeas and Stêpieñ 2008, Stêpieñ and Gazeas
2012). An important starting point of the model is based on the
observational, as well as theoretical, arguments that the orbital period
distribution of young cool binaries is concentrated around 2--3~d. This
fact is supplemented with an assumption that components of these binaries
possess magnetic winds carrying mass and AM in the same way as single stars
of the lower MS. Mass loss is rather moderate (of the order of 0.1~\MS\
during the whole MS age) so it influences little a nuclear time scale of
either component. But, as detailed computations of the evolution of cool
close binaries show, the AM loss time scale is close to MS life time of a
primary component for typical values of the initial orbital period (Stêpieñ
2011a). In effect, the typical primary reaches RLOF when it is close to
TAMS. Following RLOF, a mass transfer occurs. With dynamical effects taken
into account, it can be shown that, until mass reversal, the process takes
time comparable to the thermal time scale of the secondary component, as
opposed to the conventional models of mass transfer neglecting dynamics of
the transferred matter. SD1 variables are identified with this phase of
mass transfer. After mass ratio reversal a CB, or SD2 binary is formed,
depending on the amount of AM left in the system. Further evolution of an
SD2 binary is affected by two factors acting in the opposite direction:
mass transfer from the present secondary to the present primary makes
period increase whereas the AML makes it shorten. Depending on the relative
importance of these processes the binary evolves either into contact or
becomes an Algol (Stêpieñ 2011b).

\Acknow{We thank Wojtek Dziembowski, the referee, for a very careful
reading of the manuscript and numerous remarks which substantially improved
the original version of the paper. This research was partly supported by
the National Science Centre under the grant DEC-2011/03/B/ST9/03299.  We
acknowledge the use of the SIMBAD database, operated at CDS, Strasbourg,
France.}

\end{document}